\documentclass[preprint,journal]{vgtc}            

\onlineid{1199}

\vgtccategory{Research}

\vgtcpapertype{algorithm/technique}

\title{Peripheral Teleportation: A Rest Frame Design to Mitigate Cybersickness During Virtual Locomotion}

\author{%
  \authororcid{Tongyu Nie}{0000-0003-4186-8749},
  \authororcid{Courtney Hutton Pospick}{0000-0002-4203-1513},
  \authororcid{Ville Cantory}{0009-0004-5153-2020},
  \authororcid{Danhua Zhang}{0000-0001-5284-211X},
  \authororcid{Jasmine Joyce DeGuzman}{0009-0007-5962-145X},\\
  \authororcid{Victoria Interrante}{0000-0002-3313-6663},
  \authororcid{Isayas Berhe Adhanom}{0000-0003-4798-7415},
  and \authororcid{Evan Suma Rosenberg}{0000-0002-4826-4561}
}

\authorfooter{
  \item
        Tongyu Nie, Courtney Hutton Pospick, Ville Cantory, Danhua Zhang, Victoria Interrante, and Evan Suma Rosenberg are with the Department of Computer Science \& Engineering at University of Minnesota. 
  	E-mails: \{nie00035, hutto070, canto063, zhan5954, interran, suma\}@umn.edu
  \item
  	Jasmine Joyce DeGuzman is with the Department of Computer Science at University of Central Florida.
  	E-mail: jasdeg@ucf.edu

  \item
        Isayas Berhe Adhanom is with Department of Computer Science at Texas State University.
  	E-mail: isayas@txstate.edu
}

\abstract{
Mitigating cybersickness can improve the usability of virtual reality (VR) and increase its adoption. 
The most widely used technique, dynamic field-of-view (FOV) restriction, mitigates cybersickness by blacking out the peripheral region of the user’s FOV.
However, this approach reduces the visibility of the virtual environment. 
We propose \textit{peripheral teleportation}, a novel technique that creates a rest frame (RF) in the user's peripheral vision using content rendered from the current virtual environment. 
Specifically, the peripheral region is rendered by a pair of RF cameras whose transforms are updated by the user's physical motion.
We apply alternating teleportations during translations, or snap turns during rotations, to the RF cameras to keep them close to the current viewpoint transformation.
Consequently, the optical flow generated by RF cameras matches the user's physical motion, creating a stable peripheral view. 
In a between-subjects study (N=90), we compared peripheral teleportation with a traditional black FOV restrictor and an unrestricted control condition.
The results showed that peripheral teleportation significantly reduced discomfort and enabled participants to stay immersed in the virtual environment for a longer duration of time.
Overall, these findings suggest that peripheral teleportation is a promising technique that VR practitioners may consider adding to their cybersickness mitigation toolset.
} 


\teaser{
  \centering
  \includegraphics[width=\textwidth, trim = 0px 20px 0px 40px, clip]{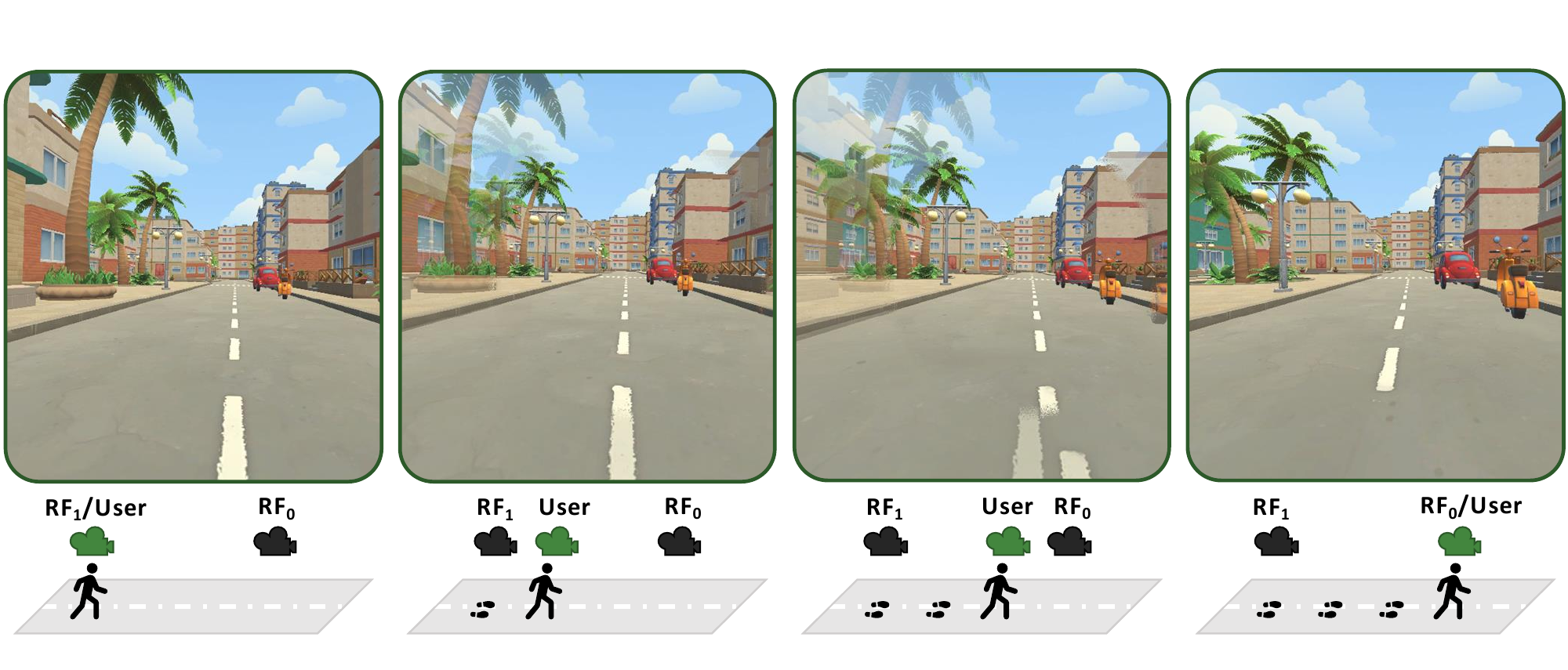}
  \caption{
  During virtual locomotion, peripheral teleportation transitions between two rest frames, represented here by the black camera icons in the bottom row as $RF_{0}$ and $RF_{1}$. 
  When virtual locomotion starts, $RF_1$ has the same transform (position and orientation) as the headset while $RF_0$'s transform is predicted based on the current virtual locomotion velocity and a pre-defined teleportation time interval $T$. 
  When the user travels between $RF_{0}$ and $RF_{1}$, the image in the periphery of the user's camera is replaced with a linear interpolation between $RF_{0}$ and $RF_{1}$.
  Additionally, both $RF_0$ and $RF_1$'s local transforms are updated in the same way as the headset's so that their rendered output matches the physical motion of the user.
  }
  \label{fig:pt_time}
}

\graphicspath{{figs/}{figures/}{pictures/}{images/}{./}} 

\usepackage{booktabs}                  

\usepackage{mathptmx}                  

\begin{document}


\firstsection{Introduction}

\maketitle

Locomotion is one of the most common tasks when interacting with a virtual environment (VE) \cite{laviola_3d_2017}. 
Real walking is more efficient \cite{suma_evaluation_2010} and provides a higher sense of presence \cite{usoh_walking_1999}, but space constraints often limit its use \cite{al_zayer_virtual_2020}. 
Continuous virtual locomotion, typically implemented using a handheld controller, enables users to navigate without these physical constraints, but is widely associated with cybersickness. 
Virtual turns, in particular, pose a greater risk of discomfort compared to virtual translations \cite{wu_asymmetric_2022, zielasko_systematic_2022}. 
Susceptibility to cybersickness varies significantly across individuals \cite{tian_least_2024}, and remains a phenomenon not fully explained by current theories \cite{tian_review_2022}. Discrete viewport control techniques, such as teleportation and snap turning, can reduce cybersickness compared to continuous virtual locomotion, but often at the cost of disorienting users \cite{bowman_travel_1997, kelly_teleporting_2020, farmani_evaluating_2020}.

The discomfort caused by continuous locomotion is often attributed to the optical flow generated by virtual motion.  Although optical flow plays an important role for path integration, it can also induce cybersickness when the virtual camera motion deviates from the user's physical movements.  Fortunately, distinct optical flow patterns emerge in different regions of the field-of-view (FOV), and the visual system processes these regions of optical flow with varying sensitivities \cite{stoffregen_flow_1985}. Peripheral optical flow plays a key role in postural control and vection, while foveal optical flow is mostly useful for perceiving direction of motion \cite{warren_direction_1988}. Based on this, previous researchers introduced dynamic field-of-view restriction, also known as \textit{tunneling}, to block peripheral optical flow and reduce cybersickness during virtual locomotion \cite{bolas_dynamic_2014, fernandes_combating_2016}.  FOV restriction has become one of the most widely implemented cybersickness mitigation techniques in virtual reality applications.

Although FOV restriction can potentially reduce discomfort, artificially reducing the field-of-view may also have negative side effects.  
For example, a wider field-of-view (FOV) 
enhances
immersion \cite{cummings_how_2015}, improves depth perception \cite{buck_comparison_2018, jones_peripheral_2013, wu_perceiving_2004}, and boosts task performance \cite{ragan_effects_2015, kim_display_2022, grinyer_effects_2022}. 
Alternative FOV manipulation techniques, such as blurring \cite{lin_how_2020, nie_analysis_2020} and contrast reduction \cite{zhao_mitigation_2022}, can potentially mitigate cybersickness without fully blocking the periphery.
However, reducing image quality in the periphery may also impair users' ability to detect and respond to peripheral stimuli \cite{duinkharjav_image_2022, lu_attributes_2014}. 
Furthermore, research has also suggested that blurring is less effective at reducing cybersickness because some of the optical flow from continuous virtual locomotion in the periphery is preserved \cite{groth_visual_2021}.

In this work, we developed \textit{peripheral teleportation}, a novel cybersickness mitigation technique that creates a rest frame (RF) in the periphery using cameras placed close to the user.
We apply consecutive teleportations and snap turns to the RF cameras to keep them close to the user's current transform and serve as a rest frame to mitigate discomfort.  These discrete movements allows meaningful visual content to be displayed in the periphery without introducing continuous optical flow that may exacerbate cybersickness.   
To evaluate this technique, we conducted a between-subject study with 90 participants that compared peripheral teleportation, a standard FOV restrictor that occludes the same peripheral area, and an unrestricted control condition.  Participants in the peripheral teleportation condition reported less discomfort and were able to stay immersed in the virtual environment significantly longer compared to both the FOV restrictor and control conditions.  These results indicate that peripheral teleportation is a promising new cybersickness technique that may be implemented as an alternative to commonly used FOV restrictors.

\section{Related Works}
\subsection{The Cause of Cybersickness}
Since Stanney et al. \cite{stanney_aftereffects_1998} identified the need to develop a better theoretical understanding of cybersickness, 
different models concerning sensory conflict, postural instability, etc., are in development, although this goal has yet to be fully achieved \cite{stanney_identifying_2020}. 
The sensory conflict theory claims that the cause of cybersickness is the discrepancy between the visual and inertial information.
The optical flow generated by virtual locomotion in VR specifies acceleration that does not correspond to the inertial force changes users can perceive \cite{lackner_motion_2014, oman_motion_1990, laviola_discussion_2000}.
On the other hand, the postural instability theory attributes cybersickness to prolonged exposure to postural instability \cite{riccio_ecological_1991}.
In VR, optical flow does not fully correspond to the vestibular/inertial information in accordance with physical laws, creating difficulties for the user in stabilizing their posture.
Researchers have had success using postural data (sometimes in combination with other physiological signals) to predict cybersickness 
\cite{islam_cybersickness_2021, cortes_eeg-based_2023}. 
Although these two hypotheses disagree with each other in many aspects, both of them agree that 
Earth referents, which means elements of VR displays that are stable relative to the Earth \cite{bailey_using_2022},
should mitigate cybersickness \cite{stanney_identifying_2020, bailey_using_2022}.

\subsection{Measuring Cybersickness}

Traditionally, researchers test the cybersickness level that a user experience with subjective questionnaires.
However, objective cybersickness measures are still under exploration.

\subsubsection{Subjective Measures}
The simulator sickness questionnaire (SSQ) is widely adopted to quantify the cybersickness elicited by virtual reality \cite{kennedy_simulator_1993}. 
Although SSQ can capture the symptoms that users experience at the end of the experiment, it does not provide information about users' discomfort level in-between the start and the end of exposure.
Repeating SSQ takes a significant amount of time and could lead to higher reported symptoms due to demanding characteristics \cite{ellis_demand_2007, bimberg_usage_2020}.
The Fast Motion Sickness Score (FMS) is an easy-to-administer, finely graded motion sickness measure where participants report their level of motion sickness on a scale from 0 (no sickness at all) to 20 (frank sickness) \cite{keshavarz_validating_2011}. 
FMS can be taken repeatedly throughout the exposure to a provocative stimulus so that the time course of motion sickness can be captured.
Fernandes and Feiner adapted the discomfort score developed by Rebenitsch and Owen, where participants were asked to rate their discomfort level from 0 (how they felt coming in) to 10 (severe discomfort) \cite{fernandes_combating_2016, rebenitsch_individual_2014}. 
Each participant's discomfort score ratings can be used to calculate their average discomfort score (ADS) and relative discomfort score (RDS).
RDS captures the participant's relative performance if they terminate early \cite{fernandes_combating_2016}, and has been widely used by researchers (e.g., \cite{cao_visually-induced_2018, wu_dont_2021, norouzi_assessing_2018}). While subjective measures are widely used in assessing cybersickness, they have several limitations that impact their reliability, including their dependence on the user's ability to judge and recall their discomfort, and their inability to capture real-time fluctuations in discomfort and sickness levels.

\subsubsection{Objective Measures}
Objective measures aim to provide implicit, real-time, and continuous data on a user's levels of discomfort and cybersickness.  
Previous studies have identified various physiological signals that can serve as objective indicators of cybersickness symptoms. 
Among these, gastric activity measured by electrogastrogram (EGG) \cite{dennison_use_2016}, 
heart rate or heart rate variability measured by electrocardiogram (ECG) or Photoplethysmogram (PPG), 
brain activity measured by electroencephalogram (EEG) or functional magnetic resonance imaging (fMRI) \cite{tian_who_2024, mimnaugh_virtual_2023}, 
electrodermal activity (EDA) \cite{dennison_use_2016},   
eye movement behavior \cite{lopes_2020},  
and postural activity data 
\cite{stoffregen_postural_1998}
have been explored as potential indicators of cybersickness.
Recent approaches have also used deep learning methods to analyze information about the visual scene, such as 3D motion flow, to estimate cybersickness symptoms \cite{zhao_2023}. 
Additionally, other approaches have integrated deep learning techniques with both physiological data and visual scene information to estimate and forecast cybersickness \cite{islam_cybersickness_2021}.
Despite their potential, objective measures are still not widely used due to their complexity, cost, and technical challenges in physiological data collection and analysis.

\subsection{Mitigating Cybersickness}
\subsubsection{Overview}
Researchers have proposed various techniques to mitigate cybersickness in VR (see Ang and Quarles \cite{ang_reduction_2023} for a review).
Most of the software-based techniques, such as the dynamic FOV restrictor, rest frame, and peripheral blurring, try to reduce peripheral optical flow.
Peripheral vision plays an important role in motion perception because the human visual system is more sensitive to motion in the peripheral retina \cite{stoffregen_flow_1985, lishman_vision_1981}, and the magnitude of optical flow is larger in the outer region of the image space \cite{gibson_parallax_1955}.

\subsubsection{Dynamic FOV Restriction}
The dynamic FOV restrictor, which reduces the FOV during virtual locomotion, is the most widely researched cybersickness mitigation technique \cite{bolas_dynamic_2014}. 
Many studies have confirmed its effectiveness in reducing cybersickness 
(e.g., \cite{fernandes_combating_2016, zhao_mitigation_2022, wu_adaptive_2022}).
Previous studies on FOV restrictors used different FOVs, ranging from $80^\circ$ \cite{fernandes_combating_2016, zhao_mitigation_2022} to $55^\circ$ \cite{wu_adaptive_2022, adhanom_effect_2020}, or even smaller \cite{bala_dynamic_2021, al_zayer_effect_2019}.
Researchers have proposed different restrictor designs, including foveated restrictor \cite{adhanom_effect_2020}, fixed shape asymmetrical restrictors \cite{wu_asymmetric_2022, wu_dont_2021}, and optical-flow-reducing restrictors \cite{wu_adaptive_2022, bala_dynamic_2021}.
Others also evaluated its effect on spatial learning \cite{adhanom_field--view_2021} and navigation performance \cite{al_zayer_effect_2019}. 

\begin{figure*}[t]
    \centering
    \includegraphics[width = \textwidth, trim = 0px 330px 610px 50px, clip]{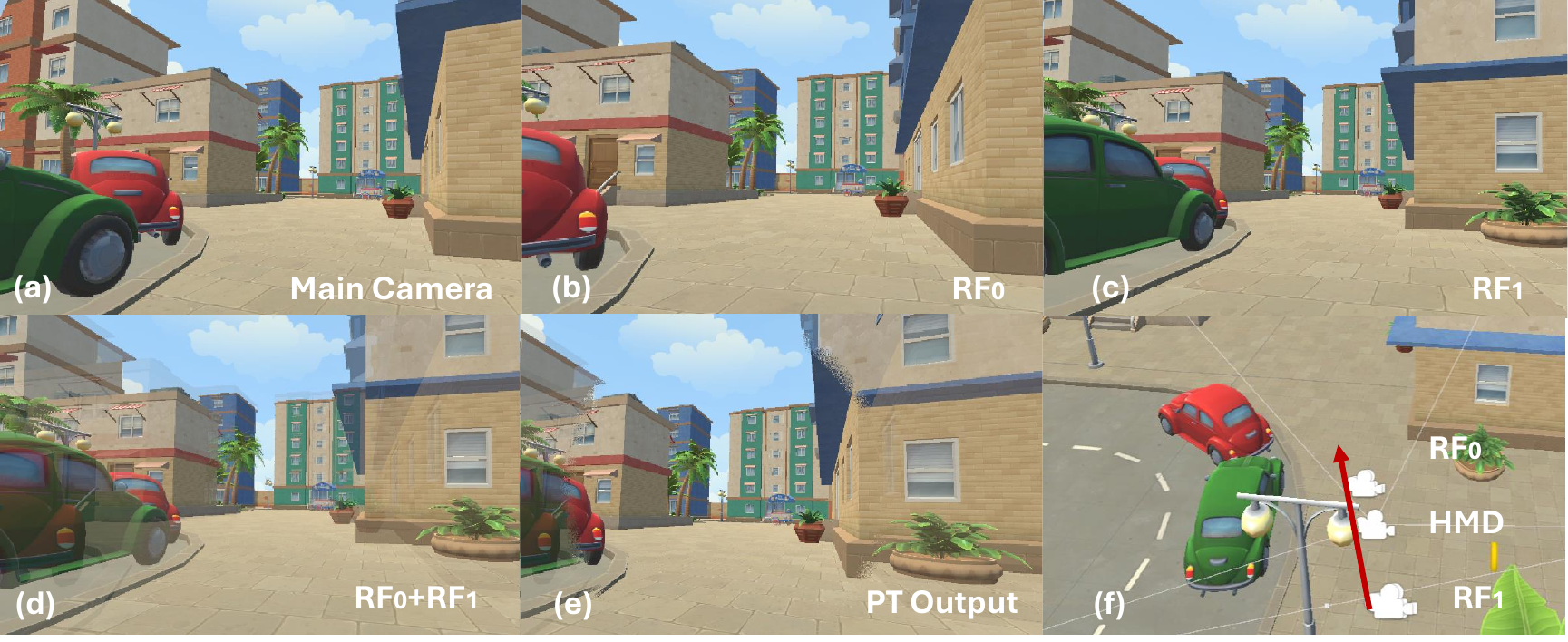}
    \caption{An overview of the peripheral teleportation technique. 
    Two extra rest frame cameras $RF_0$ and $RF_1$ were rendering images beside the main camera. 
    Their positions in the VE are illustrated in (f).
    The red arrow denotes the moving direction of the user.
    Figures (a-c) are the output of the main camera, $RF_0$, and $RF_1$.
    The rest frame (d) is a linear interpolation of images rendered by $RF_0$ and $RF_1$ (b, c).
    Finally, peripheral teleportation replaces the peripheral region of (a) with the rest frame (d)'s peripheral region, creating an output like figure (e).
    }
    \label{fig:pt_space}
\end{figure*}

\subsubsection{Rest Frames}
When the user is walking physically in VR, the artificial optical information is congruent with the user's physical motion.
In this case, people do not tend to get sick \cite{chance_locomotion_1998}. 
As a result, researchers have used rest frames, which are portions of the VE that remain fixed in relation to the real world, to mitigate cybersickness.
Most of the proposed RFs are fixed to the Earth.
A stationary visual background reduced postural disturbance induced by virtual motion \cite{duh_independent_2001}. A static metal net rest frame significantly reduced cybersickness, whereas a dynamic one did not \cite{cao_visually-induced_2018}. Researchers have also proposed other improved designs, such as adding random noise-like grains \cite{cao_granulated_2021} or a wireframe representation of the user's physical space in the periphery \cite{wu_combining_2019}, to enhance visual search task performance or ensure user safety.
However, some of the other proposed RFs are fixed to the head.
Previously, it was recognized that a virtual nose can significantly reduce cybersickness \cite{wienrich_virtual_2018}. However, a recent experiment with a larger sample size failed to replicate this finding \cite{yip_preregistered_2024}. An alternative design, called \textit{AuthenticNose}, features a more realistic appearance \cite{ang_gingervr_2020}. A comparative experiment did not show any significant effect between a head-fixed RF, FOV restriction, and blurring \cite{shi_virtual_2021}.
The Circle effect proposed by Buhler et al. in a poster shares some similarities with the proposed peripheral teleportation technique by adding a stationary camera's view into the periphery \cite{buhler_reducing_2018}.
However, they did not teleport the stationary camera and chose to blend in the main camera's image instead, which introduced undesirable optical flow.
Additionally, their user study only involved 18 participants and did not show any significant difference in cybersickness reduction.

\subsubsection{Other Techniques}
Other than the techniques mentioned above, researchers have used other solutions to mitigate cybersickness.
Software-based techniques include blurring \cite{lin_how_2020, nie_analysis_2020}, contrast reduction \cite{zhao_mitigation_2022}, reverse optical flow \cite{park_mixing_2022, buhler_reducing_2018, xiao_augmenting_2016}, path modification \cite{hu_reducing_2019}, and geometry modification \cite{groth_cybersickness_2024, nie_like_2023, lou_geometric_2022}.
Hardware-based techniques include galvanic vestibular stimulation \cite{groth_omnidirectional_2022, sra_adding_2019}, vibration \cite{peng_walkingvibe_2020, jung_floor-vibration_2021} and haptic feedback \cite{liu_phantomlegs_2019}.
Distraction \cite{venkatakrishnan_effects_2023, venkatakrishnan_effects_2024}, action \cite{lin_intentional_2022}, or heartbeat feedback \cite{joo_effects_2024} can also reduce discomfort.
However, there was a lack of a comparison between the software-based techniques that produced significant results.
Some of the techniques, like reverse flow visualization or geometry modification, can introduce undesirable visual effects in the fovea.
Additionally, the adoption of hardware-based techniques is limited because they require extra devices.

\subsection{Discrete Movement}
First mentioned by Mine \cite{mine_virtual_1995}, teleportation is a target-based VR locomotion interface that involves discrete viewpoint movement and generally does not induce cybersickness compared to continuous locomotion \cite{adhikari_integrating_2022, rahimi_scene_2020, prithul_teleportation_2021, langbehn_evaluation_2018}.
However, during teleportation, the rapid view change disoriented users \cite{bowman_travel_1997} because self-motion cues are important to spatial updating \cite{kelly_teleporting_2020}.
Researchers have proposed various solutions to improve the spatial awareness of teleportation (see Prithul et al. \cite{prithul_teleportation_2021} for a review).
\textit{Jumping} is a range-restricted variant of teleportation \cite{weissker_spatial_2018}. Pulsed interpolation shows a sequence of viewpoints along the path to serve as a middle ground between teleportation and continuous locomotion. However, it did not perform significantly better than teleportation in the spatial task \cite{rahimi_scene_2020}.
HyperJump adds teleportation to continuous locomotion, which merges benefits from both techniques and allows faster locomotion compared to continuous locomotion \cite{adhikari_integrating_2022}. 
Subconscious adjustments to the viewpoint transform during saccades or blinks are effective strategies for increasing gains during redirected walking \cite{sun_towards_2018, langbehn_blink_2018}. To enhance teleportation, techniques such as Smart Avatars and Stuttered Locomotion have been proposed, where the former provides path awareness through avatar visualization, and the latter repeatedly teleports users upon locomotion initiation \cite{freiwald_continuity_2022}. Comparative analyses of snap turning techniques revealed that selection-based methods outperform directional methods in search tasks \cite{zielasko_systematic_2022}.

\begin{figure*}
  \centering
  \includegraphics[width=\textwidth, trim = 0px 328px 165px 20px, clip]{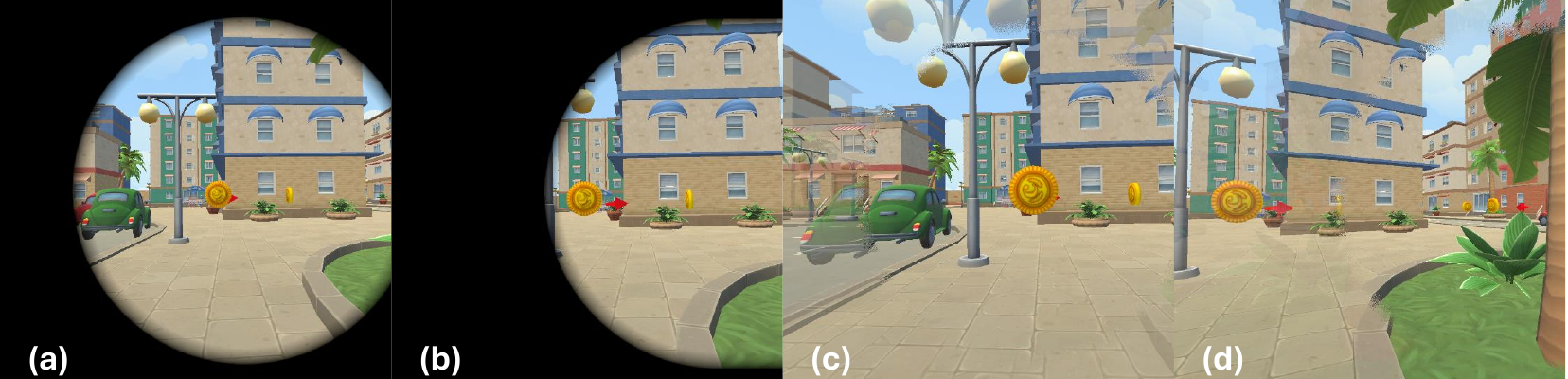}
  \caption{
    During translation, the appearance of a symmetric FOV mask in the black FOV restrictor condition (a) and the peripheral teleportation condition (c) is a circle. 
    During turning, we used an asymmetric mask that shifted its center into the turn (b, d). 
    All images were for the left eye.
  }
  \label{fig:pt_shape}
\end{figure*}

\section{Peripheral Teleportation Design}
\subsection{General Design}\label{sec:pt_design}
As shown in Figure \ref{fig:pt_space}, the peripheral teleportation technique uses the output of two additional pairs of cameras ($RF_0$ and $RF_1$) to create 
an Earth referent
in the periphery and mitigate cybersickness.
$RF_0$ and $RF_1$ each contain two cameras for stereoscopic rendering, which we refer to as a whole 
 to simplify our notation.
$RF_0$ and $RF_1$ are attached to camera rigs $Rig_0$ and $Rig_1$ that serve as the parent objects.

\subsubsection{Rest Frame Design}
For each frame, the local transforms of $RF_0$ and $RF_1$ are updated with the local transform of the main camera relative to its parent object.
As a result, the optical flow generated by both RF cameras is also congruent with the user's physical motion, which should be less likely to induce cybersickness.
Although the blending effect mixes the optical flow from two cameras together, since both RF cameras' optical flow are coupled with physical motion, their combination is also coupled with physical motion.

\subsubsection{Teleportation Design}
To keep the peripheral image similar to the main camera's view without introducing continuous virtual motion, we update $Rig_0$ and $Rig_1$'s transforms discretely.
Figure \ref{fig:pt_time} provides an overview of this process.
When the user starts virtual locomotion, $Rig_1$ is placed at the same position and orientation as the HMD's camera rig.
$Rig_0$ is teleported to the predicted transform given the main camera rig's current transform and predefined translational and rotational velocity $\nu$ and $\omega$ after $T$ seconds.
We set $T=1$ second in our implementation.
In a more generalized case, where $\nu$ and $\omega$ are not predefined, the user's current speed can always be used for prediction.

The images of RF cameras, $I_{RF0}$ and $I_{RF1}$, are dynamically blended to keep the peripheral content similar to the main camera's view.
For pixels in the periphery, their values are determined based on equation \ref{eq:pt_lerp}, where $t$ is the time since the last teleportation.
\begin{equation}\label{eq:pt_lerp}
    I_{Peri}=\mbox{lerp}(I_{RF1}, I_{RF0}, t/T) \mbox{where } t\in[0, T] 
\end{equation}
As a result, the periphery will be exactly the same as the HMD's view when $t=0$, which allows us to dynamically scale the FOV mask without delay when the user starts or stops moving. 
This design was kept the same for the FOV restrictor condition for scientific rigor.
The periphery will also be the same as the HMD's view when $t=T$ if the user does not deviate from the predicted path.
Based on equation \ref{eq:pt_lerp}, at time $t=T$, the intensity of $RF_1$ is 0, making it invisible to the user.
Since $RF_1$ is invisible, we can set $Rig_0$'s transform to $Rig_1$ and $Rig_0$'s transform to a new predicted transform without being noticed.
The value of $t$ is reset to 0 and the loop goes on until the user stops virtual locomotion.
If the user terminates the navigation when $t<T$, their FOV scales back.
Since the peripheral effect is not visible at this moment, all parameters are reset to their initial state.

\subsection{Dithering Transition} \label{sec:dither}
\begin{figure}[t]
    \centering
    \includegraphics[width = \columnwidth, trim = 0px 320px 400px 0px, clip]{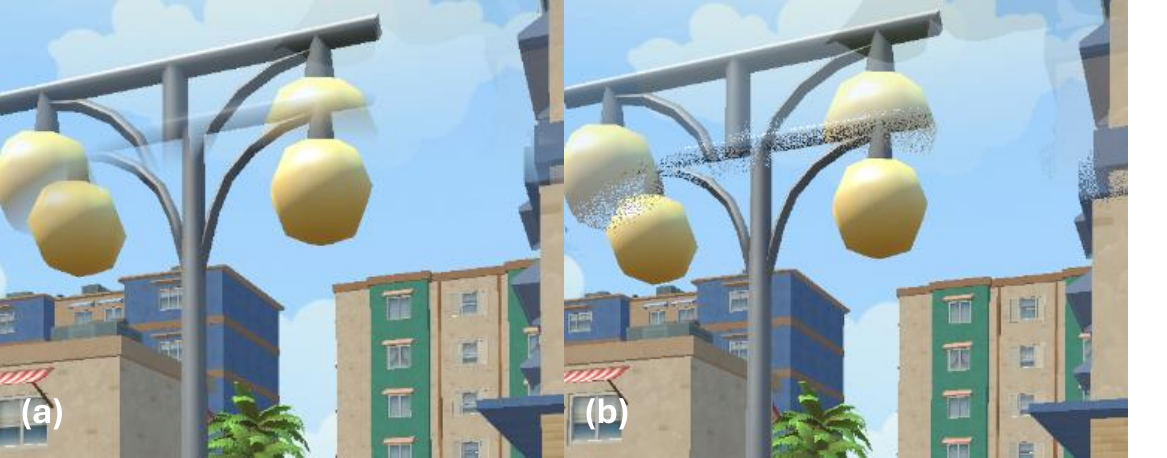}
    \caption{
    The alpha blending (a) and dithering (b) transition effect between the fovea and the periphery.
    For the dithering effect, the probability of a pixel between the inner and outer radius of the restrictor showing the image of the rest frame decreases from 1 to 0 as it moves from the outer radius to the inner radius.
    }
    \label{fig:dither}
\end{figure}

Noise can provide high frequency spatial detail and enhance the image quality \cite{tariq_noise-based_2022}. 
During development, we had initially implemented peripheral teleportation using linear interpolation between the two RF camera images.  However, this temporal transition effect was more noticeable than we had originally expected, especially during turns.  After experimenting with different transitions, we found that a ``dissolve'' effect, also known as dithering, was less visually intrusive.  
Figure \ref{fig:dither} illustrates the difference between linear interpolation and dithering for a single frame during a transition. 
For pixels between the outer and inner radius $r_{out} $ and $ r_{in}$, the probability of the pixel showing the peripheral content, $p$, is defined in equation \ref{eq:prob}.
$r_{px}$ is the pixel's corresponding FOV.
For both the black FOV restrictor and peripheral teleportation, $r_{out} $ and $ r_{in}$ were set to $60^\circ$ and $55^\circ$ in the user study.
\begin{equation}\label{eq:prob}
    p = \frac{r_{px} - r_{in}}{r_{out} - r_{in}}
\end{equation}
For each pixel, we generate a pseudo-random number $n_r\in[0, 1]$.
The final color value of the pixel is randomly determined based on equation \ref{eq:dither}. 
\begin{equation}\label{eq:dither}
    I_{trans}= \left\{ \begin{array}{rcl}
         I_{peri} & \mbox{if} & n_r < p \\
         I_{fovea} & \mbox{if} & n_r > p 
    \end{array}\right.
\end{equation}

\subsection{Asymmetric Lateral Shape}
An asymmetric lateral FOV restrictor outperforms a traditional symmetrical restrictor in mitigating cybersickness and improving environment visibility \cite{wu_asymmetric_2022}.
As a result, we adopted the same FOV restrictor design where the foveat region's center is shifted $17^\circ$ towards the turning direction during turning. 
This design is kept the same in our user study for the black restrictor and the peripheral teleportation except for the transition effect mentioned in section \ref{sec:dither}.
Figure \ref{fig:pt_shape} shows the shape of the side restrictor during translation and turning for both the black restrictor and peripheral teleportation.

\subsection{Implementation and Performance}
We implemented peripheral teleportation in Unity with a compute shader. 
We created four cameras using the same parameters as the main camera.
Two were put under $Rig_0$ and form $RF_0$ while the other two were under $Rig_1$ and form $RF_1$.
Each camera's target texture was set to a render texture.
For every frame update, we synthesized the peripheral view by executing a compute shader.
The shader took all render textures from the main and RF cameras as input, and output the image with a rest frame in the periphery using the method described in sections \ref{sec:pt_design}.
We adopted the way of calculating FOV from VR Tunnelling Pro\footnote{\url{https://github.com/sigtrapgames/VrTunnellingPro-Unity}}.

We measured the GPU time for our demo with peripheral teleportation on and off during locomotion using PIX\footnote{\url{https://devblogs.microsoft.com/pix}}.
On average, peripheral teleportation increased GPU time from $1.64$ ms to $3.53$ ms.
Since running at $72$ Hz allows for $13.88$ ms latency, the increased GPU time is acceptable.

\section{User Study}
\subsection{Experiment Design}
We conducted a between-subjects study to evaluate the proposed technique with three conditions:
\begin{itemize}
    \item Peripheral teleportation (PT)
    \item Black FOV restrictor (Restrictor)
    \item Standard viewpoint with full FOV (Control)
\end{itemize}
We chose a between-subjects design because previous studies have reported order effects in within-subject studies \cite{fernandes_combating_2016, cao_visually-induced_2018}.
This experiment was conducted in our lab following a study protocol that was reviewed and approved by the Institutional Review Board (IRB) at the University of Minnesota.

\subsection{Participants}
A total of 90 participants (45 male, 45 female) participated in the study.
Participants' ages ranged from 19 to 27 years old (M = 23.57, SD = 2.67).
They were recruited from classroom announcements, email lists and flyers posted around the campus.
We required participants to be over the age of 18, able to stand without assistance, have normal or corrected vision, without history of severe motion sickness, and not pregnant.
They were randomly assigned to three groups with a balanced gender distribution.
Participants' self-reported video game experience is listed in Table \ref{table:game_exp}.
We compensated each participant with extra credit or a \$15 Amazon gift card.

\subsection{Equipment}
The experiment was conducted with a Meta Quest 2 headset and Oculus Link. 
The headset has a resolution of $1832 \times 1920$ per eye, a $97^\circ$ FOV, and a refresh rate of 72 Hz.
The VR application was running on a PC with an Intel Core i9-9900K CPU, an NVIDIA RTX 2080 Ti GPU, and 32GB RAM.

\begin{table}[]
\begin{tabular}{llll}
                         & Restrictor & Control & PT \\
I don't play video games & 7          & 13      & 14 \\
Less than 1 year         & 3          & 0       & 2  \\
2-4 years                & 3          & 0       & 1  \\
5-9 years                & 3          & 3       & 1  \\
10 or more years         & 14         & 14      & 12
\end{tabular}
\caption{
    Participants' gaming experience categorized by condition.
    }
    \label{table:game_exp}
\end{table}

\subsection{Virtual Environment}
\begin{figure}[t]
    \centering
    \includegraphics[width = \columnwidth, trim = 0px 20px 0px 60px, clip]{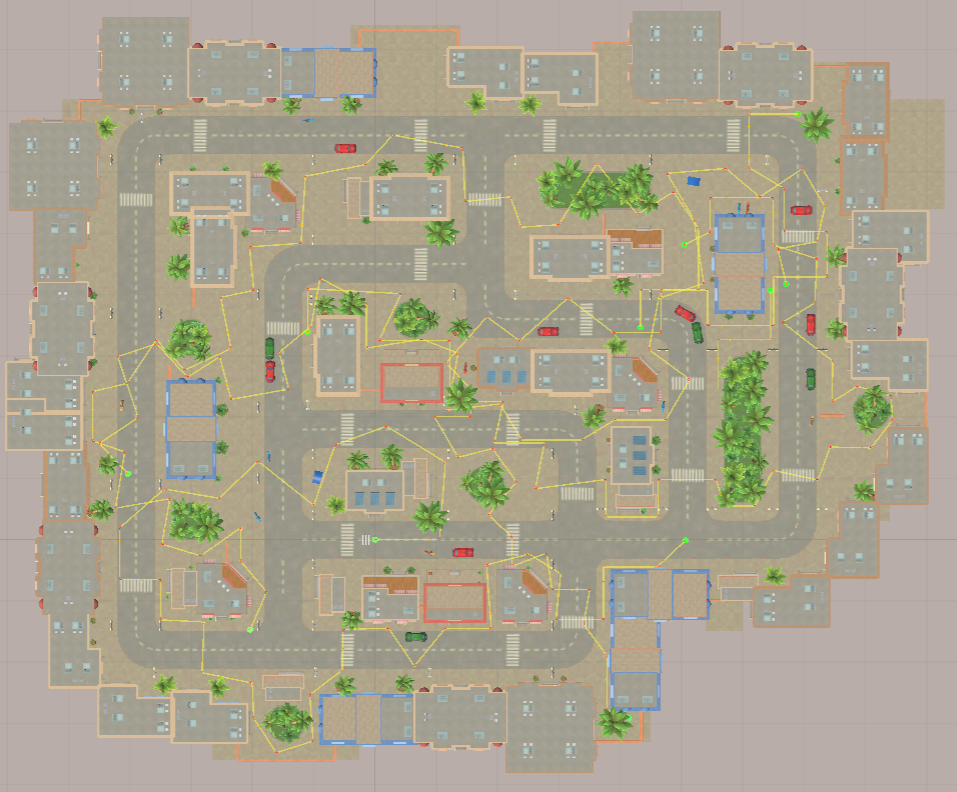}
    \caption{
    An overhead view of the virtual environment. 
    The environment was about $130 m\times 100 m$ in size.
    Participants followed paths defined by coins and arrows in different trials, which were visualized in yellow lines.
    }
    \label{fig:VE}
\end{figure}

The experiment task required participants to navigate a city
(see Figure \ref{fig:VE})
while wearing a head-mounted display (HMD).
The VE was built using an 
asset\footnote{\url{https://assetstore.unity.com/packages/3d/environments/urban/modular-resort-town-91427}} 
from the Unity asset store.
The VE included 10 different paths, each corresponding to a trial.
Each path was defined by a set of gold coins and red arrows and took about 2.5 minutes to navigate.
Each path required participants to make 35 virtual turns.
We implemented the virtual locomotion interface using view-directed steering with the XR Interaction Toolkit package 1.0.0-pre.8 in Unity.
The maximum translational velocity was $3 m/s$.
The maximum rotational velocity was $45^\circ/s$.
The VE was implemented in Unity 2021.3.23f1.

\subsection{Procedure}
The whole experiment took about 45 minutes to complete, including all questionnaires. 
At the beginning of the study session, the experimenter received confirmation that the participant met all of the inclusion criteria, introduced the task, showed the participant how to use the controller, and reviewed the IRB-approved information sheet.
Before the participant entered the virtual space, they completed the SSQ pre-questionnaire on a PC.
Participants were asked to report their discomfort level at the end of each trial and stop the VR experience if they were feeling too sick to continue.

Participants remained standing during the VR experience. 
They could virtually translate and turn using the controller via the thumbstick on an Oculus Touch controller using view-directed steering.
We instructed participants to use the joystick to turn.
They can move their head but they are instructed to avoid turning their whole body with their feet.
All paths contain consecutive turns in the same direction where participants cannot negotiate by only turning their head.
There was a short practice trial at the beginning of the VR experience to familiarize participants with the task.
Ten experimental trials follow the practice trial, which were designed to take $2.5$ minutes each, resulting in an overall VR exposure time of about $25$ minutes. 
For each trial, they were given one of the predefined virtual paths that passed through the VE.
The order of trials was randomly shuffled for every participant.
At the end of a trial, participants were asked to rate their discomfort level on a 0 to 10 scale using a slider before they went to the next trial.  
The VR experience terminated when the participant finished all the trials or whenever they indicated they would like to stop using VR by taking off the headset or entering a discomfort score of 10 at the end of any trial.
In the post-study questionnaire, participants completed the SSQ post-questionnaire, the iGroup presence questionnaire \cite{schubert_experience_2001}, a subjective feedback questionnaire, and a demographic questionnaire. 

\subsection{Measures}
\subsubsection{Simulator Sickness}
We collected participants' SSQ responses before and after their VR experience and calculated the changes between their responses.
We did not analyze the difference between the SSQ sub-scores because we do not have specific hypotheses about them.
We also asked participants ``are you motion sick right now?'' at the beginning of the SSQ.

\subsubsection{Discomfort Score}
We collected discomfort scores after every trial by asking the participants to self-report their level of discomfort using a slider in VR.
We adopted the \textit{Average Discomfort Score (ADS)} and the \textit{Relative Discomfort Score (RDS)} proposed by Fernandes and Feiner \cite{fernandes_combating_2016} using the following equations:
\begin{equation}
    ADS = \frac{\sum\limits_{0\leq i \leq t_{stop}}DS_i}{N} 
\end{equation}
\begin{equation}
    RDS = \frac{\sum\limits_{0\leq i \leq t_{stop}} DS_i + (t_{max}-t_{stop}+1)DS_{stop}}{t_{max}}
\end{equation}
The VR experience duration for each participant was $t_{stop}$. 
The longest duration among all participants was $t_{max}$. 
The last discomfort score rating at $t_{stop}$ was $DS_{stop}$. 
If a participant terminated before $t_{max}$, we set their $DS_{stop}$ to $10$ and repeated it for all subsequent time. 

\subsubsection{Task Duration}
We measured the time each participant spent on the navigation task because the locomotion task terminated when they were feeling severe discomfort. 
For each trial, the VR application recorded the time between when they started moving and when they stopped at the end of the trial.
We summed the duration of each trial to get the participant's task duration.

\subsubsection{Visibility and Presence}
We asked the participant to rate agreement with the statement ``It was difficult to see the virtual environment during locomotion.'' on a 7-point Likert scale, which we refer to as \textit{visibility} in our results.
We used the Igroup Presence Questionnaire (IPQ) \cite{schubert_experience_2001} to measure the sense of presence in VR. 
It has 3 subscales: spatial presence, involvement and experienced realism, and an additional general item which assesses the general "sense of being there". 
\subsection{Hypotheses}
We defined the following scientific hypotheses to evaluate the effects of peripheral teleportation.
The hypotheses were preregistered on Open Science Framework \footnote{\url{https://osf.io/b3mkd/?view_only=94699921c3e24cc195bd9778b939809e}}.
\begin{itemize}
    \item \textbf{H1}: Participants would report lower delta SSQ scores with the peripheral teleportation and FOV restrictors compared to the control condition without restriction. 
    \item \textbf{H2}: Participants would report lower discomfort scores with the peripheral teleportation and FOV restrictors compared to the control condition.
    \item \textbf{H3}: Participants would be immersed in the navigation task longer with the peripheral teleportation and FOV restrictors compared to the control condition. 
    \item \textbf{H4}: Participants would report better visibility with the peripheral teleportation and control condition compared to the FOV restrictor.
    \item \textbf{H5}: Participants would report a greater sense of presence with the peripheral teleportation and control condition compared to the FOV restrictor.
\end{itemize}


\begin{figure}
    \centering
    \includegraphics[width = \columnwidth, trim = 0px 5px 0px 0px, clip]{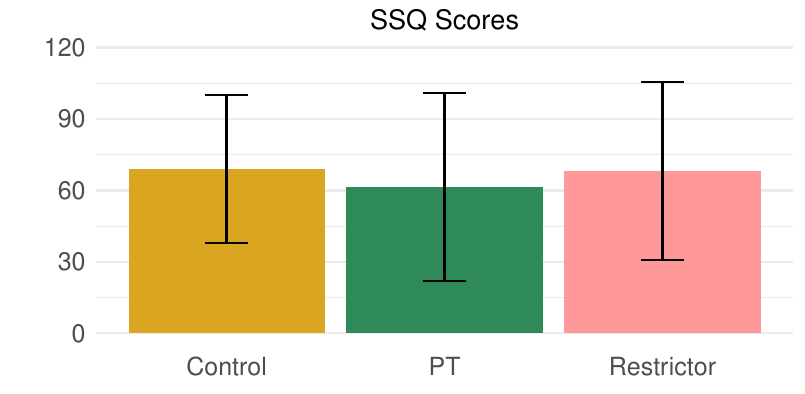}
    \caption{
    Bar charts with the mean and standard deviation of SSQ scores. 
    There were no significant differences between conditions.
    }
    \label{fig:ssq}
\end{figure}

\begin{figure}
    \centering
    \includegraphics[width = \columnwidth, trim = 0px 5px 0px 0px, clip]{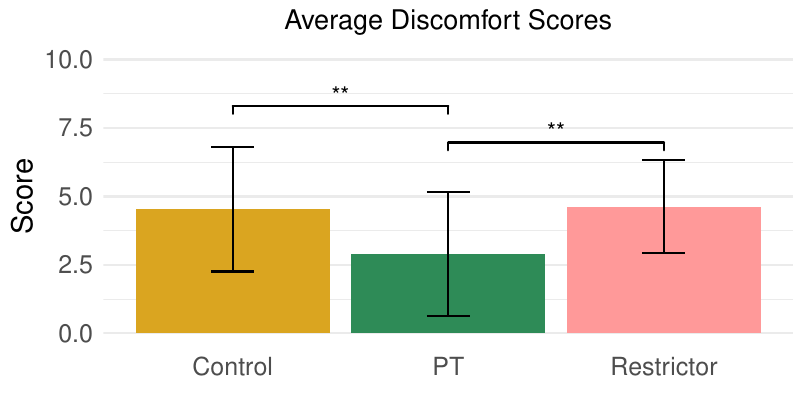}
    \caption{
    Bar charts show mean and standard deviation of average discomfort scores (ADS). 
    A value of 0 means ``no discomfort at all'' throughout all trials.
    The greater the score is, the more discomfort users experienced.
    Users reported significantly lower ADS in the PT condition than the  Restrictor and Control conditions.
    Note: Throughout all figures, significance levels are indicated as follows: * $p<.05$, ** $p<.01$, *** $p<.001$.
    }
    \label{fig:ads}
\end{figure}

\section{Results}
We first conducted Shapiro-Wilk tests for all variables. 
The SSQ, ADS, and the involvement and realism subscales of IPQ data did not show significant violation of normality, while the rest did. 
For parametric data, we used between-subjects ANOVAs to analyze differences between the three conditions (Restrictor, Control, PT) and reported descriptive statistics as mean (\textit{M}) and standard deviation (\textit{SD}).
For the non-parametric data, we used Kruskal-Wallis tests and reported descriptive statistics as median (\textit{Mdn}) and interquartile range (\textit{IQR}). 
Statistical tests assumed a significance value of $\alpha=.05$. 
When a Kruskal-Wallis test rejected the null hypothesis, we conducted the post hoc analysis using Conover tests with a Holm correction.
We reported Cliff's $\delta$ as the effect size for non-parametric data, where $\delta>.474$ indicates a large effect size.

\subsection{Simulator Sickness Questionnaire}
For the question "are you motion sick right now?", 26 participants answered `Yes' in Control, 21 in Restrictor, and 19 in PT.
Results of delta SSQ scores did not violate the normality assumption and 
are
shown in Figure \ref{fig:ssq}. 
Analysis results of the differences between the pre- and post-SSQ scores indicated no significant difference among the 
Control ($M=69.07$, $SD=31.06$), 
Restrictor ($M=68.07$, $SD=37.42$) and 
PT conditions ($M=61.34$, $SD=39.47$), $F(2)=0.37$, $p=.69$. 
This result does not support H1.

\subsection{Discomfort Scores}
\subsubsection{Average Discomfort Scores}
Figure \ref{fig:ads} shows the ADS results.
The analysis for ADS showed significant differences between the three conditions, $F(2)=6.18$, $p=.004$, $\eta^2=.130$. 
Post hoc comparisons found the PT condition ($M=2.89$, $SD=2.26$) was significantly more comfortable compared 
with the Restrictor condition ($M=4.63$, $SD=1.70$), $p=.006$,
Cohen's $d=.83$,
and the Control condition ($M=4.53$, $SD=2.28$), $p=.006$,
$d=.78$.
The Restrictor and Control's ADS ratings were not significantly different, $p=.86$,
$d=.05$.
These results partially support hypothesis H2.

\subsubsection{Relative Discomfort Scores}
\begin{figure}[t]
    \centering
    \includegraphics[width = \columnwidth, trim = 0px 20px 0px 0px, clip]{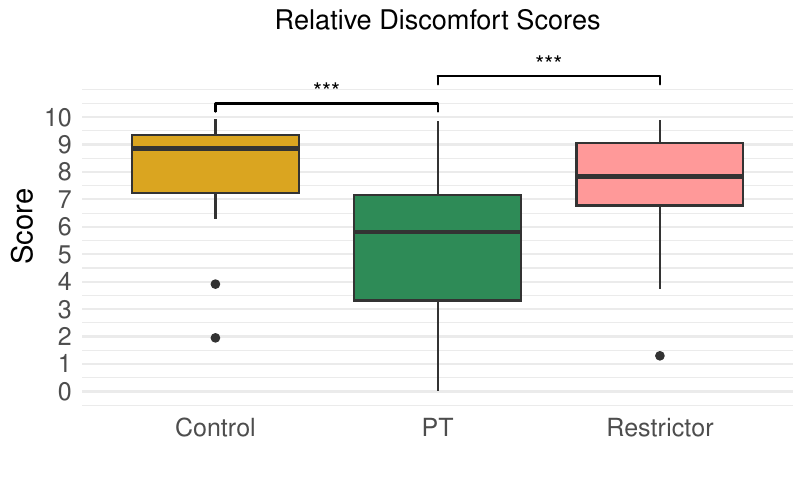}
    \caption{
    Box plots of relative discomfort scores (RDS).
    Users reported significantly lower RDS in the PT condition than in the Restrictor and Control conditions.
    }
    \label{fig:rds}
\end{figure}

Figure \ref{fig:rds} shows the RDS results.
For RDS, the analysis was significant as well, $\chi^2(2)=22.18$, $p<.001$.
Post hoc comparisons indicated the PT condition's RDS ($Mdn=5.80$, $IQR=3.84$) was significantly lower 
than Control's ($Mdn=8.85$, $IQR=2.13$), $p<.001$,
$\delta=.64$,
and Restrictor's ($Mdn=7.83$, $IQR=2.27$), $p<.001$,
$\delta=.55$. 
The Restrictor and Control's RDS ratings were not significantly different, $p=.25$,
$\delta=.18$.
These results partially support hypothesis H2.

\subsection{Task Duration}\label{sec:duration}
\begin{figure}[t]
    \centering
    \includegraphics[width = \columnwidth, trim = 0px 20px 0px 0px, clip]{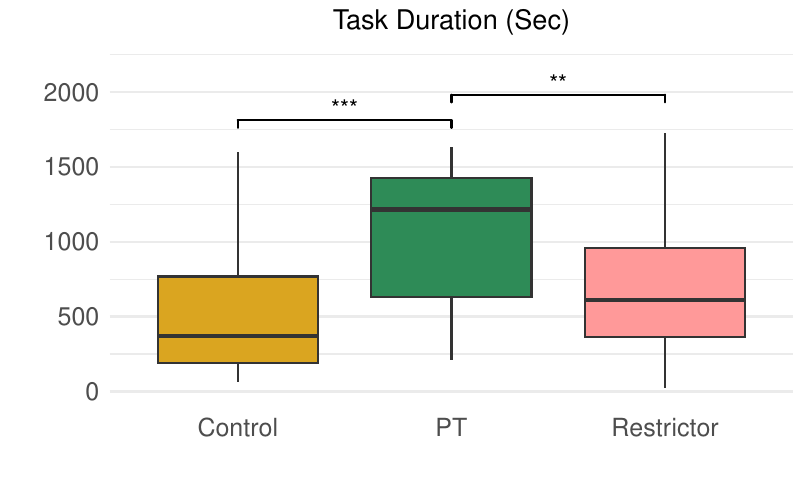}
    \caption{
        Box plots of task duration for each condition.
        Participants lasted significantly longer in the PT condition than in the Restrictor and Control conditions.
    }
    \label{fig:duration}
\end{figure}

Figure \ref{fig:duration} shows the results (sec) for overall task duration.
The analysis on task duration revealed significant differences among the three conditions, $\chi^2(2)=17.52$, $p<.001$.
Post hoc comparisons found the PT condition's task duration ($Mdn=1215$, $IQR=795.25$) was significantly longer 
than Restrictor's ($Mdn=609.5$, $IQR=592.75$), $p=.005$,
$\delta=.46$,
and Control's ($Mdn=371$, $IQR=576.25$), $p<.001$,
$\delta=.58$.
The Restrictor and Control conditions' task duration was not significantly different, $p=.15$, 
$\delta=.23$.

\subsection{Visibility}
The visibility scores (Restrictor: $Mdn=5$, $IQR=3$, Control: $Mdn=6$, $IQR=3$, PT: $Mdn=6$, $IQR=1.75$) were not significantly different, $\chi^2(2)=1.63$, $p=.44$.


\subsection{Presence}
\begin{figure*}[t]
    \centering
    \includegraphics[width = \textwidth, trim = 0px 20px 0px 20px, clip]{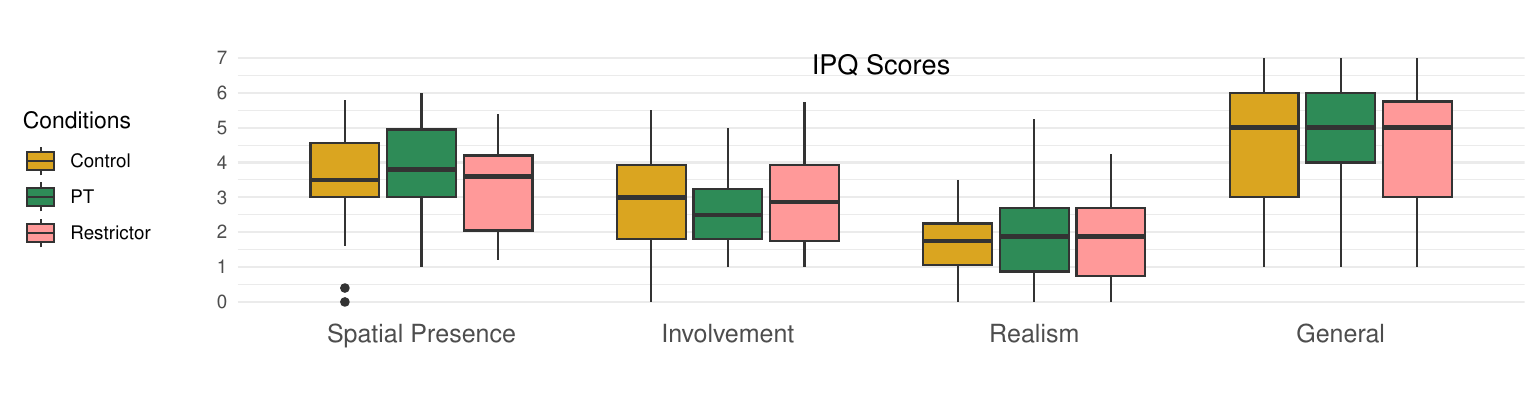}
    \caption{
    Results for the Igroup Presence Questionnaire.
    There were no significant differences between conditions for all subscales.
    }
    \label{fig:ipq-results}
\end{figure*}

Figure \ref{fig:ipq-results} shows the results for all IPQ sub-scales.
Between-subject ANOVAs showed no significant difference among the three conditions for the involvement sub-scale, $F(2)=.76$, $p=.47$, and realism sub-scale $F(2)=.32$, $p=.73$.
Kruskal-Wallis tests showed no significant difference for the general sub-scale, $\chi^2(2)=1.00$, $p=.61$, and spatial presence sub-scale, $\chi^2(2)=1.81$, $p=.40$.

\section{Discussion}
\subsection{Cybersickness and Task Duration}

Although the SSQ scores were not significantly different, this can likely be attributed to the large differences in immersion time between the conditions.
To minimize harm to participants, the study protocol required stopping the experience when they became motion sick or entered a discomfort score of 10, and the similarity in delta SSQ scores across the conditions suggests that cybersickness severity was fairly consistent when a stop criterion was reached.
Previous studies of FOV restriction, including the original work Fernandes and Feiner, observed similar results with this type of study protocol (e.g., \cite{fernandes_combating_2016, wu_dont_2021, cao_visually-induced_2018}).  
Our work is therefore consistent with previous research, which supports the conclusion that discomfort scores collected throughout the experiment provide a more sensitive measurement when consistent immersion time is not enforced.

In the peripheral teleportation condition, the median immersion time was over 20 minutes, which was approximately double the amount of time compared to the FOV restrictor and over three times as long as the control condition.  
The navigation scenario was designed to take around 25 minutes to complete, and so most participants in the PT condition were fairly close to finishing the task.
Therefore, even though the delta SSQ scores were not significantly different between conditions, the large effect size we observed for task duration was strong evidence that PT was effective in mitigating cybersickness.

\subsection{Discomfort Scores}
The results for both average and relative discomfort scores also support the effectiveness of peripheral teleportation.  We had originally hypothesized that both peripheral teleportation and FOV restriction would provide benefits over the control condition.  However, we did not expect to observe significant differences between the two cybersickness mitigation techniques, because both conditions were implemented with consistent parameters that occluded the same peripheral region during virtual locomotion.  However, participants in the peripheral teleportation condition also reported significantly lower discomfort compared to the FOV restrictor, which exceeded our expectations.  These findings are of particular interest to the research field for the following three reasons.

First, the results confirmed the effectiveness of a dynamic rest frame.
Traditionally, the rest frame technique has been considered to be a weaker mitigation technique, because most prior studies have failed to produce significant results \cite{ang_reduction_2023}.
Despite this, ecological psychologists have predicted that rest frames could be among the most effective cybersickness mitigation techniques due to their ability to provide optical information that is coupled with the user's physical motion \cite{bailey_using_2022}.
Our results support this prediction, which suggests that the lack of significant findings in prior studies may be limited to specific experimental conditions and should not be overgeneralized to the all rest frame techniques.

Second, despite its widespread use, Ang and Quarles have pointed out that very few studies have compared FOV restriction with other cybersickness mitigation techniques \cite{ang_reduction_2023}.
Specifically, previous studies have not directly 
compared an
FOV restrictor
with an Earth referent that has a consistent occluded area.
To the best of our knowledge, our work is the first to systematically evaluate these two techniques together in a controlled study conducted over a reasonably large sample size.


Finally, peripheral teleportation is the first software-based cybersickness mitigation technique that has shown significant improvements over the commonly used FOV restrictor.
Many earlier papers that have proposed novel mitigation techniques did not compare them with FOV restriction (e.g., \cite{cao_visually-induced_2018, park_mixing_2022, lin_how_2020, lin_intentional_2022, nie_analysis_2020, peng_walkingvibe_2020, groth_cybersickness_2024, joo_effects_2024, venkatakrishnan_effects_2023}).
Additionally, previous work that has proposed novel variants of black, symmetric FOV restriction, such as foveated or asymmetric restrictors, have not found significant improvements in cybersickness mitigation \cite{zhao_mitigation_2022, wu_dont_2021, wu_asymmetric_2022, wu_adaptive_2022, adhanom_effect_2020, bala_dynamic_2021}, although in some cases other benefits were noted such as better environment visibility. However, in this study, peripheral teleportation was more effective than the FOV restrictor; therefore, these results open up new directions for cybersickness research, especially rest frame techniques that remain underexplored.

\subsection{Visibility and Presence}
We did not observe any significant effect on either the Visibility question or IPQ responses.
Therefore, we cannot draw any scientific conclusions at this time for H4 and H5.
Even so, this result is congruent with Wu and Suma Rosenberg's original paper, in which a side restrictor had already achieved similar visibility and presence ratings compared to their control condition \cite{wu_asymmetric_2022}.
We are potentially seeing a ceiling effect here; for this reason further studies are needed to evaluate the effect of peripheral teleportation in areas other than cybersickness.

\subsection{Qualitative Feedback}

Although the study was primarily quantitative, additional insight can be gained from participants' comments on the feedback post-questionnaire.  In both the PT and FOV restriction conditions, participants mentioned that the peripheral visual effect helped to reduce discomfort.  Some participants also 
specifically
mentioned that the lateral shift during turns was helpful because the ``narrowing field-of-view corresponded with which direction I was facing.''

In the peripheral teleportation condition, some participants still found the visual effect to be noticeable despite the peripheral image's similarity to the main camera's image.
This indicates that the peripheral teleportation effect is not completely invisible, similar to other rest frame techniques.
Some participants described the shifting perspective or the overlapping nature of the image as ``blurry'' or ``wasn't consistent.''  
However, these visual artifacts may be more noticeable during turning, because it was commonly stated that PT helped 
lessen
discomfort when moving in a straight direction.
Some participant comments also suggest that they perceived the visual effect as a rest frame, such as ``the screen follows my head.''  Only one participant mentioned that the peripheral region was distracting.

In the FOV restrictor condition, some participants also found the visual effect to be undesirable, with comments such as ``disorienting'' and ``irritable on the eyes.''  A couple participants described the changing field-of-view size as ``odd" or ``out of place,'' and one participant even went as far as to call it ``oppressive.''

Taken together, these qualitative comments suggest that software-based cybersickness mitigation techniques often involve a tradeoff between physiological discomfort and other aspects of the subjective user experience.  We suggest that the ideal parameters for rest frame and FOV restriction techniques may vary between individuals, based on their susceptibility to cybersickness and personal preferences.

\subsection{Limitations and Future Work}

As with any experiment, there are limitations in generalizability that should be acknowledged.  Although we recruited a reasonably large sample size with equal numbers of men and women, the age of participants skewed younger than the general population because we primarily recruited from the University campus.
We also evaluated the cybersickness mitigation techniques in a specific type of virtual environment, a large outdoor urban scene.
Peripheral teleportation and FOV restriction may offer varying benefits and tradeoffs in environments that produce different locomotion behavior, such as building interiors that require more close-quarter maneuvering.
We believe that systematically comparing cybersickness mitigation techniques using a diverse sampling of representative virtual environments would be valuable in future work.  
Future studies could also investigate the effectiveness of these techniques with respect to individual differences in video game experience and cybersickness susceptibility.

This paper presents the initial design and evaluation for peripheral teleportation, and future work could make improvements upon this technique.  
Currently, the peripheral teleportation technique may introduce discontinuities in the visual content within the FOV, resulting in a visual artifact similar to screen tearing at the restrictor mask boundary.
We believe that a gaze-contingent rest frame design using an eye-tracked VR headset would have strong potential to make any undesirable visual artifacts from peripheral teleportation less noticeable.
Similarly, a more sophisticated trajectory prediction method for the RF cameras could potentially enhance the subjective user experience.

We selected parameters for virtual locomotion and peripheral effects based on informal testing in the lab during the design and development of the experiment.  However, it would also be valuable to more comprehensively evaluate the effects of various parameters in future work.  For example, the teleportation interval $T$ was set to $1$ sec to make sure the teleportation is perceived as discrete, while some researchers suggest it can be decreased to $0.3$ sec \cite{wolwer_how_2024}.  A smaller $T$ could make the peripheral visual effect less noticeable, because it would be updated more frequently.  Additionally, the effectiveness and intrusiveness for both PT and FOV restriction can depend upon the shape and size of the peripheral region where the visual effect is applied.
Moreover, using quantitative measures, such as a two-alternative forced choice task or decoy questions, can be helpful for measuring user tolerance to deviations in their expected trajectory. 

The effect of peripheral teleportation cannot be attributed solely to optical flow.
Specifically, the observed effect may partly arise from attention shifting, as peripheral vision processes changes that diverge from typical experiences.
Distractions can reduce the perceived severity of cybersickness by diverting users’ conscious awareness from the symptoms \cite{venkatakrishnan_effects_2023}.
Similarly, changes in the periphery may act as visual distractions, thereby decreasing reported cybersickness symptoms. 
Consequently, side effects of peripheral teleportation, such as attention shifting or luminance changes, should be analyzed independently to gain a comprehensive understanding of why this technique is effective.
Lastly, we believe that individual variability in user preferences and customizable cybersickness mitigation interfaces would be worthwhile topics for future investigation.

\section{Conclusion}
In this work, we proposed \textit{peripheral teleportation}, a novel rest frame design for mitigating cybersickness in virtual reality experiences that use continuous virtual locomotion.
We conducted a between-subjects study that compared peripheral teleportation, a black FOV restrictor, and a control condition.
When using PT, participants reported less discomfort and were able to stay immersed in the virtual environment for a significantly longer duration of time.
These results suggest that rest frame techniques such as peripheral teleportation may be more promising than previously reported in the literature. 
In conclusion, peripheral teleportation represents a new method that virtual reality practitioners may consider adding to their cybersickness mitigation toolset and a valuable candidate for future studies that can evaluate the benefits and tradeoffs these techniques.

\acknowledgments{
The authors wish to thank Sahar Aseeri, John Schroeder, Seraphina Yong, Yeena Ng, Serena Xin, The Council of Graduate Students at University of Minnesota, and other people we do not know their names for their help with recruiting participants.
The authors also wish to thank Ryan P. McMahan for his insights on the discussion section and the anonymous reviewers for their feedback.
This work was supported by the National Science Foundation under Grant No. 1901423.
}

\bibliographystyle{abbrv-doi-hyperref}

\bibliography{template}
\end{document}